\documentclass[preprint,12pt]{elsarticle}
\usepackage{amsmath}
\usepackage{amssymb}
\usepackage[dvipdfm,colorlinks]{hyperref}
\usepackage{multirow}
\usepackage{enumitem}

\biboptions{compress}

\begin{document}

\begin{frontmatter}

\title{Tripartite Blind Quantum Computation}

%\author[label1]{}
\author{Min Liang}\ead{liangmin07@mails.ucas.ac.cn}
\address{Data Communication Science and Technology Research Institute, Beijing 100191, China}

\begin{abstract}
This paper proposes a model of tripartite blind quantum computation (TBQC), in which three independent participants hold different resources and accomplish a computational task through cooperation. The three participants are called C,S,T separately, where C needs to compute on his private data, and T has the required quantum algorithm, and S provides sufficient quantum computational resources. Then two concrete TBQC protocols are constructed. The first protocol is designed based on Broadbent-Fitzsimons-Kashefi protocol, and it cannot prevent from collusive attack of two participants. Then based on universal quantum circuit, we present the second protocol which can prevent from collusive attack. In the latter protocol, for each appearance of $R$-gate in the circuit, one call to a classical AND-BOX is required for privacy.
\end{abstract}

\begin{keyword}
quantum cryptography \sep blind quantum computation \sep computing on encrypted data \sep universal quantum circuit \sep secure two-party computation
\end{keyword}

\end{frontmatter}

% \linenumbers
\section{Introduction}
Blind quantum computation (BQC) \cite{childs2005,vedral2012} is a practical quantum information technique combining quantum cryptography and quantum computation. It is mainly used in the cases that one party has limited quantum computational resources, and delegates his quantum computation to the other party who can do it. For example, the quantum computer is so expensive for personal purchasing, however, the corporation or organization can purchase the quantum computer and make money by providing quantum computing services to individual customers. Then there is a problem: the customers do not trust the corporation, and worry that his private data, result and even the quantum algorithm may be learned by the corporation. The problem can be solved using BQC protocol, by which the quantum computer can provide secure cloud services.

BQC deals with the problem: one party with limited computational power delegates his computational task to another party, and can get the required result while his privacy (the private data and result) is secured during the whole process. This problem is firstly studied by Childs \cite{childs2005}, and then lots of BQC protocols have been proposed \cite{aharonov2008,arrighi2006,broadbent2009,sueki2012,morimae2010,morimae2012a,morimae2012b,dunjko2012,rohde2012,fisher2013}. There are also some development about the experiment of BQC \cite{fisher2013,barz2012,barz2013}. In current studies about BQC, the data is supposed to be provided by Client, and the computational service is supposed to be provided by Server, but the algorithm may be provided by Client or Server or be public. So, the researches of BQC protocols can be classified into three kinds. (1) The algorithm is private and belongs to Client. Most of researches are of this kind \cite{childs2005,aharonov2008,broadbent2009,sueki2012,morimae2010,morimae2012a,morimae2012b}; (2) The algorithm is private and belongs to Server, see Ref.\cite{rohde2012}; (3) The algorithm is public or at least agreed on by both Server and Client \cite{arrighi2006,fisher2013}. No matter which party the algorithm belongs to, the owner of it should provide better algorithm to improve the efficiency.

Because the algorithm is an important information, the owner of it wishes to keep it private during the process of BQC. Currently, the studies focus on the case that the private algorithm belongs to Client or Server \cite{childs2005,broadbent2009,rohde2012}. However, there exists an ordinary case in practice: the algorithm belongs to the third party, which intends to make money by providing the algorithm in a private way. In this case, the algorithm can be kept private while making money, so this mode can encourage the researchers to concentrate on the studies of algorithm and improve the algorithm.

This paper considers BQC from the aspect that the data, algorithm and computational resources belong to three independent parties. The model of tripartite blind quantum computation (TBQC) is proposed, and the possible adversaries are analyzed, and then three properties such as correctness, blindness, verifiability are defined. Then, based on Broadbent-Fitzsimons-Kashefi protocol \cite{broadbent2009}, a TBQC protocol is presented. Finally, the other TBQC protocol is proposed based on universal quantum circuit (UQC).

\section{TBQC model}
\subsection{Parties and protocol}
The TBQC model involves three independent participants: Client, Server and Third party, denoted as C,S,T, where C holds the data, S holds the quantum computational resources, and T holds the quantum algorithm. T has the ability of designing quantum algorithm, and prevents it from revealing during the interactive process of TBQC. In order to provide better quantum computing services, S tries to provide more quantum computational resources or stronger universal quantum computer. C has limited quantum computational power and no efficient quantum algorithm, but he intends to compute on some private data with quantum computer. He hopes to accomplish the computational task by using T's quantum algorithm and S's quantum computational resources, and prevents his data and the required result from revealing during the computation.

In the TBQC model, the three parties C,S,T play different roles in the computation. They interactively perform some local (classical or quantum) computation and communicate through classical or quantum channel, and finally finish a common computational task while the privacy is secured. The detailed interactive process is called TBQC protocol. Here the TBQC protocol is denoted as $\pi(x,f)$, where two parameters $x,f$ must be given as the inputs. The symbol $x$ is the private data owned by C, and the symbol $f$ is a mapping $f:x\mapsto f(x)$, which represents a computational task. At the beginning of the protocol, C would inform T about the computational task $f(*)$, then T can design a quantum algorithm accordingly. The encoding of the algorithm is denoted as $\mathcal{A}_f$. Given an input data $x_0$, the output of the protocol $\pi$ can be described as $\pi(x_0,f)$. If $\pi(x_0,f)=f(x_0)$, we say the protocol is correct. The definition of correctness is as follows.

{\bf Definition 1 (Correctness)}. A tripartite blind quantum computation protocol $\pi$ is correct if, given any input data $x_0$ and any required computational task $f:x\mapsto f(x)$, its outcome $\pi(x_0,f)$ equals to $f(x_0)$ when the three parties T, C and S follow the steps of the protocol honestly.

\subsection{Adversarial model}
In TBQC model, the privacy contains the data, algorithm and result. C must prevent his data and result from revealing, and T must prevent his algorithm from revealing. According to the different attacking targets, the adversarial model can be classified into three kinds: attacking the data, attacking the algorithm, and attacking the result.

The attacker may be some one of the three parties or two of them. The outer hacker may also participate the attacking. However, the outer hacker cannot obtain more information about the privacy than any one of the three parties. If the protocol can prevent inner attackers, then it can also prevent outer attackers. Thus, when considering adversarial model, it is analyzed only the inner attack by the dishonest participants.

During the interaction of the protocol, T sends his quantum algorithm as messages in a private way, so the attackers or dishonest parties may take some measures to obtain some information about the algorithm. Because the algorithm is private information that belongs to T, T must provide the algorithm in a private way, and then the other parties (C or S) can only use the algorithm, but cannot obtain the algorithm during the computation.

Similarly, in order to secure the data, C has to provide the data in a private way (the most common way is encryption), and then the other parties can only use them in computation, but cannot learn them.

Attacking the result includes two aspects: (1) the passive adversaries do not disturb the execution of protocol, but snatch some information about the final result; (2) the active adversaries maliciously disturb the execution of the protocol, and cause a wrong outcome.

In order to prevent the result from passive adversaries, it is required that the final result can only be decrypted by C in the protocol. The malicious disturbance may be caused by either the outer attackers or the inner participants. For instance, T provides an incorrect quantum algorithm, or S does not obey the protocol strictly. For this kind of active adversaries, when designing a TBQC protocol, a mechanism should be introduced to detect the destruction, so that C can judge the correctness of the final result. The disturbance may also be caused by outer attackers who break the communication channel. This kind of attack exists usually in all communication protocol, and will not be considered in this paper.

Besides, there exists another kind of adversaries: in order to attack the privacy of certain party, the other two parties may cooperate by sharing their information with each other. For example, S and T may cooperate in order to attack the data and result, which is the privacy of C. This kind of adversaries is called collusive attack.

\subsection{Security definitions}
Because the data, the algorithm, and the result are private information, a secure TBQC protocol must prevent these information from revealing. This property is called blindness. Its definition is as follow.

{\bf Definition 2 (Blindness)}. A TBQC protocol $\pi$ is blind if, given any input data $x_0$ and any required computational task $f: x\mapsto f(x)$, nothing about $x_0$ and the quantum algorithm $\mathcal{A}_f$ of $f$ and the outcome $\pi(x_0,f)$ is leaked. Here the quantum algorithm $\mathcal{A}_f$ is a series of instructions of quantum operators, and belongs to T; The input $x_0$ and the outcome $\pi(x_0,f)$ belong to C.

Because the protocol may be disturbed by the attackers and does not executed as expected, C must judge the correctness of the final result. Thus, a verifiable mechanism is required to check the final result. It is another property of TBQC protocol, and is called verifiability.

In the following, the honestly executed TBQC protocol is denoted as $\pi$, while the actually executed protocol is denoted as $\pi'$. So, the protocol $\pi'$ may be different with ideal protocol $\pi$, where there exists adversaries.

It is worth to notice that, there exists some problems that are hard to solved, however, if given a result, it is easy to verify whether it is right or not. We take the prime factorization problem for example. It can be described simply as a mapping $f:x\mapsto(p,q)$, where $x=p\cdot q$. If the protocol $\pi$ has not been disturbed, it outputs the result $\pi(x,f)=(p,q)$, where $p\cdot q=x$, otherwise, it outputs $\pi'(x,f)=(p',q')$, where $p'\cdot q'\neq x$. Obviously, the right result $(p,q)$ and the wrong one $(p',q')$ can be distinguished with certainty. After the TBQC protocol is executed with the input $x,f$, the Client can verify the output $(p',q')$: if $p'\cdot q'\neq x$, then the protocol $\pi$ is not executed as expected, and the Client rejects the result.

It can be seen from the example that, for a class of problems such as NP problems, the answers can be verified easily. Then the TBQC protocol for computing them is verifiable. However, this kind of verifiability is not universal for all problems. So it is called weak verifiability here. The definition is as follow.

{\bf Definition 3 (Weak Verifiability)}. We say a TBQC protocol $\pi$ is verifiable for a class $\mathcal{C}$ of problems if, given any input data $x_0$ and any $f\in\mathcal{C}$, the two outcomes $\pi(x_0,f)$ and $\pi'(x_0,f)$ can be distinguished with probability more than $\frac{1}{2}$.

For a TBQC protocol with weak verifiability, it is verifiable for a class of computational problems $\mathcal{C}$, but not verifiable for other problems. If a verifiable mechanism is added into the TBQC protocol, the verifiability becomes independent of the problems \cite{fitzsimons2012}. We define this kind of verifiability as strong verifiability.

{\bf Definition 4 (Strong Verifiability)}. We say a TBQC protocol $\pi$ is strong verifiable, if it is verifiable for all computational problems.

\section{TBQC based on measurement-based quantum computation}
Ref.\cite{broadbent2009} proposed a BQC protocol. We can modify it and get a TBQC protocol. In the protocol, some classical and quantum information should be transferred privately with other quantum cryptographic primitives. Because the focus of this paper is TBQC protocol, we will not state how to secure the communication between them.

Before the execution of TBQC protocol, C should inform T about the computational task, then T designs a quantum algorithm accordingly. Next, the first TBQC protocol is presented as follow. We adapt the same notations as Ref.\cite{broadbent2009}.

\begin{description}
  \item[Protocol 1]
  \item[1. C encrypts data:] For each input column $(x=0,y=1,\cdots,m)$ corresponding to C's data, (1) C randomly selects $\theta_{0,y}\in\{\frac{k\pi}{4}|k=0,1,\cdots,7\}$, and applies $Z_{0,y}(\theta_{0,y})$; (2) C randomly selects $i_{0,y}\in\{0,1\}$, and applies $X^{i_{0,y}}_{0,y}$. C sends the qubits to S in a secure way. C sends all the $\theta_{0,y}$ and $i_{0,y}$ to T in a secure way.
  \item[2. T prepares qubits:] For each column $x=1,\cdots,n-1$, and each row $y=1,\cdots,m$, T randomly selects $\theta_{x,y}\in\{\frac{k\pi}{4}|k=0,1,\cdots,7\}$, and produces single qubit $|\phi_{x,y}\rangle=|+_{\theta_{x,y}}\rangle$. T sends all the qubits to S.
  \item[3. S prepares brickwork state:] After receiving all qubits from C, S informs C. After receiving all qubits from T and C, S produces some qubits $|\phi_{n,y}\rangle=|+\rangle(y=1,\cdots,m)$. Using all the received qubits and prepared qubits, S creates an entangled state according to their indices, by applying Controlled-Z gates between the qubits in order to create a brickwork state.
  \item[4. T and S compute interactively:] For each column $x=0,1,\cdots,n-1$, and each row $y=1,\cdots,m$, according to the algorithm designed by T, T and S carries out some interaction and measurement (the details can be seen in Ref.\cite{broadbent2009}). Finally, T obtains some measured values $s_{x,y}$ (these values are unknown for S), and S can obtain $m$ qubits, which is just the encrypted result that is required by C.
  \item[5. Secure communication:] T sends all the measured values $s_{x,y}$ to C in a secure way; S sends the final $m$ qubits to C in a secure way.
  \item[6. Decryption:] After receiving the $m$ qubits and all the values $s_{x,y}$, C applies Pauli operation $Z^{s^Z_{n,y}}X^{s^X_{n,y}} (y=1,\cdots,m)$ on all the $m$ qubits, where the values of $s^Z_{n,y}$ and $s^X_{n,y}$ can be calculated from the values $s_{x,y}$.
\end{description}

Protocol 1 is analyzed from the correctness and security (blindness and verifiability) in the following.

{\bf Theorem 1 (Correctness)}. Assume C, S and T follow the steps of Protocol 1, then the outcome is correct.

{\it Proof}. This protocol is modified from the protocol in Ref.\cite{broadbent2009}. In fact, this modification changes only the way of interaction, but has not changed the computing process. Obviously, if the participants T and S join together, Protocol 1 would degenerate into the protocol in Ref.\cite{broadbent2009}. Thus, the outcome of this protocol is also correct. $\hfill{}\Box$

{\bf Theorem 2 (Blindness)}. Protocol 1 is blind.

{\it Proof}. The blindness of the protocol should be analyzed from the following three points. The former two points hold in the case that S and T would not cooperate. (1) The data will not leak. C encrypts his data and sends the encrypted data and the key to S and T separately. If S and T do not cooperate, the data will not leak. (2) The result will not leak. After the interaction of S and T, T owns only the decryption key, and S owns only the encrypted result. If they do not cooperate, the result will not leak. (3) The algorithm will not leak. In the protocol, the algorithm is only used in the interaction of T and S. Because the protocol in Ref.\cite{broadbent2009} does not reveal the algorithm, Protocol 1 does not leak  the algorithm. In conclusion, Protocol 1 is blind.$\hfill{}\Box$

Protocol 1 does not satisfy strong verifiability. It only satisfies weak verifiability, that means the obtained result can be verified for only a class of problems.

{\bf Theorem 3}. Protocol 1 cannot defeat collusive attack of S and T.

{\it Proof}. In the first step, C sends the encrypted data and the key to S and T separately; So S and T can get the data if they cooperate. In the fourth step, S and T can get the encrypted result and the key separately; So they can also get the result if they cooperate. Thus, Protocol 1 cannot defeat collusive attack of S and T.$\hfill{}\Box$

In the next section, the second TBQC protocol is proposed, and it can defeat collusive attack.

\section{TBQC based on universal quantum circuit}
\subsection{Main idea}
{\bf Definition 5 (Universal Quantum Circuit \cite{bera2010})}. Fix $n>0$ and let $\mathcal{U}$ be a collection of unitary transformations on $n$ qubits. A quantum circuit $C_{\mathcal{U}}$ on $n+m$ qubits is universal for $\mathcal{U}$ if, for each transformation $U\in\mathcal{U}$, there is a string $e_U\in\{0,1\}^m$ (the encoding) such that for all strings $d\in\{0,1\}^n$ (the data),
\begin{equation}
C_\mathcal{U}(|d\rangle\otimes|e_U\rangle)=(U|d\rangle)\otimes|e_U\rangle.
\end{equation}

According to the definition, we can propose a TBQC protocol based on UQC. The main idea is as follows: the data $|d\rangle$ is provided by C, the encoding $|e_U\rangle$ of the transformation $U\in\mathcal{U}$ is provided by T, and the UQC (or quantum computer) $C_\mathcal{U}$ is provided by S. By using the encoding provided by T and the quantum circuit provided by S, C can finish the quantum information processing on his data $|d\rangle$, and obtain the result $U|d\rangle$.

For a given computational task, T designs a quantum algorithm based on the quantum transformations in the set $\mathcal{U}$, that is finding a series of quantum transformations in $\mathcal{U}$, for example $U_1,U_2,\cdots,U_v\in\mathcal{U}$, such that performing these transformations in order can implement the given computational task. The encoding of the algorithm is the combination of the encodings of the series of quantum transformations $U_i(i=1,\cdots,v)$, denoted as $e_i$. Using the UQC $C_\mathcal{U}$, the computation of the algorithm $U_1,U_2,\cdots,U_v$ is carried out as follows:
\begin{eqnarray}
&& C_\mathcal{U}(|d\rangle\otimes|e_1\rangle)=(U_1|d\rangle)\otimes|e_1\rangle,\nonumber\\
&& C_\mathcal{U}((U_1|d\rangle)\otimes|e_2\rangle)=(U_2U_1|d\rangle)\otimes|e_2\rangle,\nonumber\\
&& \cdots\cdots\cdots\cdots \nonumber\\
&& C_\mathcal{U}((U_{v-1}\cdots U_1|d\rangle)\otimes|e_v\rangle)=(U_v\cdots U_2U_1|d\rangle)\otimes|e_v\rangle.
\end{eqnarray}
Finally, the result $U_v\cdots U_2U_1|d\rangle$ is obtained.

From the above analysis, we can see that, based on UQC, the three parties T,C,S can finish some quantum information processing on any data when they cooperate together. However, their cooperation should be designed carefully, in order to protect the privacy of T and C. Next, we design a secure way of cooperation, and get a TBQC protocol.

If the owner S of the UQC is treated as the distrusted third party, then the construction of TBQC protocol based on UQC is similar to designing a protocol for secure two-party quantum computation with distrusted third party. The distrusted third party may obey the protocol for secure two-party quantum computation, meanwhile, he collects some information about the privacy of T and C during the computation. In addition, he may also not obey the protocol and perform some incorrect computation. Inspired by the protocol for secure two-party quantum computation in Ref.\cite{dupuis2010}, we design a TBQC protocol based on the above universal scheme of quantum computation.

The construction of UQC and the universal quantum gates are public. According to the basic quantum transformations that are implemented by the UQC, T designs the quantum algorithm, which is a series of the basic quantum transformations. Here we assume the set of quantum gates used in the construction of UQC is $\{X,Y,Z,\mathrm{CNOT},H,P,R\}$, which is the same as that in Ref.\cite{dupuis2010}. In the TBQC protocol, initially, C and T select the random numbers $Key^0_C=(C^0_x,C^0_z)$ and $Key^0_T=(T^0_x,T^0_z)$ as their initial keys, which are used to encrypt their privacy (the data $|d\rangle$ or the encoding of algorithm $|e\rangle$). Then, C and T send the ciphertext of their privacy to S who carries out the computation of UQC on the ciphertext. At the same time, C and T update their initial keys $Key^0_C,Key^0_T$ according to the order of different gates in the UQC (suppose there are $k$ quantum gates), and finally obtain the final keys $Key^k_C=(C^k_x,C^k_z),Key^k_T=(T^k_x,T^k_z)$. S carries out the UQC on the encrypted data and the encrypted encoding, and obtains an encrypted result.

In order to explain the key point clearly, we express it using some identities. First of all, some notations are introduced as follows. Suppose the data is represented by $n$ qubits, the encoding of the algorithm is a series of $m$-qubit string. The initial keys $C^0_x,C^0_z,T^0_x,T^0_z$ are four $n+m$-bit random numbers, and each of them can be regarded as two parts, where the first part has $n$ bits and the second part has $m$ bits. Let the set of indices $c=\{1,2,\cdots,n\}$ and $t=\{n+1,n+2,\cdots,n+m\}$. Later we denote $C^0_x(w),w\in c\cup t$ as the $w$-th bit of the $n+m$-bit random number $C^0_x$, and denote $C^0_x(c)=(C^0_x(1),C^0_x(2),\cdots,C^0_x(n))$, $C^0_x(t)=(C^0_x(n+1),C^0_x(n+2),\cdots,C^0_x(n+m))$. Then we have $C^0_x=(C^0_x(c),C^0_x(t))$, $C^0_z=(C^0_z(c),C^0_z(t))$, $T^0_x=(T^0_x(c),T^0_x(t))$, $T^0_z=(T^0_z(c),T^0_z(t))$. Here the $n+m$-bit numbers $C^0_x$ and $C^0_z$ are randomly selected by C, and the $n+m$-bit numbers $T^0_x$ and $T^0_z$ are randomly selected by T. For a $n$-bit number $b=b_1b_2\cdots b_n$ ($b_i\in\{0,1\}$) and a unitary operator $U$, denote $U^b=U^{b_1}\otimes U^{b_2}\otimes\cdots\otimes U^{b_n}$. The key point can be expressed as follows:
\begin{eqnarray}\label{eq1}
&& C_{\mathcal{U}}(X^{C^0_x(c)\oplus T^0_x(c)}Z^{C^0_z(c)\oplus T^0_z(c)}|d\rangle\otimes X^{C^0_x(t)\oplus T^0_x(t)}Z^{C^0_z(t)\oplus T^0_z(t)}|e\rangle) \nonumber\\
&=& X^{C^k_x(c)\oplus T^k_x(c)}Z^{C^k_z(c)\oplus T^k_z(c)}\otimes X^{C^k_x(t)\oplus T^k_x(t)}Z^{C^k_z(t)\oplus T^k_z(t)} C_{\mathcal{U}}(|d\rangle\otimes |e\rangle)\nonumber\\
&=& X^{C^k_x(c)\oplus T^k_x(c)}Z^{C^k_z(c)\oplus T^k_z(c)}\otimes X^{C^k_x(t)\oplus T^k_x(t)}Z^{C^k_z(t)\oplus T^k_z(t)} (U|d\rangle\otimes |e\rangle)\nonumber\\
&=& X^{C^k_x(c)\oplus T^k_x(c)}Z^{C^k_z(c)\oplus T^k_z(c)}(U|d\rangle)\otimes X^{C^k_x(t)\oplus T^k_x(t)}Z^{C^k_z(t)\oplus T^k_z(t)}|e\rangle.
\end{eqnarray}
From the Eq.(\ref{eq1}), we should find a UQC $C_{\mathcal{U}}$ and a key-updating algorithm that satisfy the above equations. The key-updating algorithm depends on the UQC in this scheme. This kind of construction can be blind, since S cannot obtain the data, the encoding, and the result, however, C can decrypt the encrypted result with the help of T while T cannot obtain the result.

\subsection{Protocol}
The detail of the TBQC protocol is as follows. At the beginning, C has informed T about the computational task, and T has designed the algorithm, which is a series of $m$-qubit string (suppose there are $v$ $m$-qubit strings: $e_1,\cdots,e_v$). The protocol should run $v$ times. When the protocol runs for the $i$-th time, the outcome of the $i-1$-th time and each $e_i$ is used as its input.

\begin{description}
  \item[Protocol 2]
  \item[1. Encryption of the data:] According to the first part of the random numbers $C^0_x$ and $C^0_z$, C performs the following encryption transformation on the data $|d\rangle$:
      $$|d\rangle\rightarrow X^{C^0_x(c)}Z^{C^0_z(c)}|d\rangle,$$ and sends the ciphertext to T; According to the first part of the random numbers $T^0_x$ and $T^0_z$, T encrypts the ciphertext as follow:
      $$X^{C^0_x(c)}Z^{C^0_z(c)}|d\rangle\rightarrow X^{T^0_x(c)}Z^{T^0_z(c)}X^{C^0_x(c)}Z^{C^0_z(c)}|d\rangle,$$
      and sends the new ciphertext to S;
  \item[2. Encryption of the encoding:] According to the second part of the random numbers $T^0_x$ and $T^0_z$, T performs the following encryption transformation on the encoding $|e\rangle$ of a basic transformation:
      $$|e\rangle\rightarrow X^{T^0_x(t)}Z^{T^0_z(t)}|e\rangle,$$ and sends the ciphertext to C; According to the second part of the random numbers $C^0_x$ and $C^0_z$, C encrypts the ciphertext as follow:
      $$X^{T^0_x(t)}Z^{T^0_z(t)}|e\rangle\rightarrow X^{C^0_x(t)}Z^{C^0_z(t)}X^{T^0_x(t)}Z^{T^0_z(t)}|e\rangle,$$
      and sends the new ciphertext to S;
  \item[3. Computing through UQC:] For the encrypted data and encoding, S performs the UQC on them by orderly carrying out each gate $G_j\in\{X,Y,Z,H,P,\mathrm{CNOT},R\}(j=1,\cdots,k)$ of the UQC. Finally, the obtained result is still being encrypted. S sends the encrypted result to T. (There is an exception here: when $G_j=R$ is performed on a wire $w\in c\cup t$, the wire $w$ must be transferred to C; C picks random bits $r$ and $r'$, and applies the operator $X^rZ^{r'}P^{C^{j-1}_x(w)}$ and sends the resulting quantum state to T; T picks random bits $s$ and $s'$, and applies the operator $X^sZ^{s'}P^{T^{j-1}_x(w)}$ and sends the resulting quantum state to S; Then S continues next quantum gate $G_{j+1}$.)
  \item[4. Key-updating:] C and T update their keys according to the gates and their order in the UQC: suppose the $j$-th gate in the UQC is $G_j$, C and T update their keys according to the gate $G_j$,
      \begin{eqnarray}\label{eq2}
      && Key^{j-1}_C=(C^{j-1}_x,C^{j-1}_z)\rightarrow Key^{j}_C=(C^{j}_x,C^{j}_z),\nonumber\\
      && Key^{j-1}_T=(T^{j-1}_x,T^{j-1}_z)\rightarrow Key^{j}_T=(T^{j}_x,T^{j}_z).
      \end{eqnarray}
      For a given UQC, the gates and their order are deterministic, so the key-updating algorithm is a deterministic algorithm. Later we will present a detail description about the key-updating algorithm for each gate $G_j\in\{X,Y,Z,\mathrm{CNOT},H,P,R\}$.
  \item[5. Decryption of the result:] After receiving the encrypted result (denote it as $|\phi\rangle$) from S, T performs the following decryption transformation according to his key $T^k_x(c),T^k_z(c)$:
      $$|\phi\rangle\rightarrow|\phi'\rangle=X^{T^k_x(c)}Z^{T^k_z(c)}|\phi\rangle,$$
      and sends it to C; C decrypts it according to his key $C^k_x(c),C^k_z(c)$:
      $$|\phi'\rangle\rightarrow|\phi''\rangle=X^{C^k_x(c)}Z^{C^k_z(c)}|\phi'\rangle,$$
      and gets the result finally.
\end{description}

Especially, in Ref.\cite{liang2011}, a UQC for near-trivial transformations has been constructed. The near-trivial transformations are a class of basic transformations, and any unitary transformation can be decomposed into the product of a series of near-trivial transformations \cite{bernstein1997}. Then we can design a TBQC protocol based on the UQC for near-trivial transformations. In this protocol, T can design his quantum algorithm using a series of near-trivial transformations. Though this UQC is formed by combining the gates selected from the set $\{R_z(\pm\frac{\pi}{2^j}),R_y(\pm\frac{\pi}{2^j}),j\in\mathbf{N}\}$ and generalized $\mathrm{CNOT}$s, it can be easily transformed into a new UQC consisting of the gates in $\{X,Y,Z,\mathrm{CNOT},H,P,R\}$. So we can always assume the UQC consists of the gates in $\{X,Y,Z,\mathrm{CNOT},H,P,R\}$.

\subsection{Key-updating}
In key-updating algorithm, each step of updating in Eq.(\ref{eq2}) should satisfy the following identities (ignoring an irrelevant global phase). For any single-qubit gate $G_j\in\{X,Y,Z,H,P\}$ (the gate $R$ is an exception) applying on the $w$-th qubit, the keys $C^{j-1}_x,C^{j-1}_z,T^{j-1}_x,T^{j-1}_z$ and $C^{j}_x,C^{j}_z,T^{j}_x,T^{j}_z$ satisfy
\begin{equation}
G_j(X^{C^{j-1}_x(w)\oplus T^{j-1}_x(w)}Z^{C^{j-1}_z(w)\oplus T^{j-1}_z(w)})=(X^{C^{j}_x(w)\oplus T^{j}_x(w)}Z^{C^{j}_z(w)\oplus T^{j}_z(w)})G_j;
\end{equation}
For the two-qubit gate $G_j=CNOT$ applying on the $w$-th and $w'$-th qubits, the keys $C^{j-1}_x,C^{j-1}_z,T^{j-1}_x,T^{j-1}_z$ and $C^{j}_x,C^{j}_z,T^{j}_x,T^{j}_z$ satisfy
\begin{eqnarray}
&& G_j(X^{C^{j-1}_x(w)\oplus T^{j-1}_x(w)}Z^{C^{j-1}_z(w)\oplus T^{j-1}_z(w)}\otimes X^{C^{j-1}_x(w')\oplus T^{j-1}_x(w')}Z^{C^{j-1}_z(w')\oplus T^{j-1}_z(w')}) \nonumber\\
&& =(X^{C^{j}_x(w)\oplus T^{j}_x(w)}Z^{C^{j}_z(w)\oplus T^{j}_z(w)}\otimes X^{C^{j}_x(w')\oplus T^{j}_x(w')}Z^{C^{j}_z(w')\oplus T^{j}_z(w')})G_j.
\end{eqnarray}

Next, we introduce the key-updating algorithm for each different quantum gate $G_j\in\{X,Y,Z,\mathrm{CNOT},H,P,R\}$. This algorithm here is similar to that in Ref.\cite{dupuis2010} except the two-qubit gate $\mathrm{CNOT}$. To keep the consistence of the notations, we restate the key-updating algorithm for the gates $X,Y,Z,H,P,R$ using our notations.
\begin{itemize}[leftmargin=0cm]
\item Let $G_j\in\{X,Y,Z\}$ be the Pauli gate to be executed on wire $w\in c\cup t$. We have:
  \begin{eqnarray}
  && G_j(X^{C^{j-1}_x(w)\oplus T^{j-1}_x(w)}Z^{C^{j-1}_z(w)\oplus T^{j-1}_z(w)})\nonumber\\
  &=& (X^{C^{j-1}_x(w)\oplus T^{j-1}_x(w)}Z^{C^{j-1}_z(w)\oplus T^{j-1}_z(w)})G_j,
  \end{eqnarray}
  where an irrelevant global phase factor is ignored. Thus, C updates his key as follow:
  $$C^j_x(w)=C^{j-1}_x(w),C^j_z(w)=C^{j-1}_z(w),$$
  and T updates his key as follow:
  $$T^j_x(w)=T^{j-1}_x(w),T^j_z(w)=T^{j-1}_z(w).$$
\item Let $G_j=H$ be the Hadamard gate to be executed on wire $w\in c\cup t$. We have:
 \begin{eqnarray}
  && H(X^{C^{j-1}_x(w)\oplus T^{j-1}_x(w)}Z^{C^{j-1}_z(w)\oplus T^{j-1}_z(w)})\nonumber\\
  &=& (X^{C^{j-1}_z(w)\oplus T^{j-1}_z(w)}Z^{C^{j-1}_x(w)\oplus T^{j-1}_x(w)})H,
  \end{eqnarray}
  where an irrelevant global phase factor is ignored. Thus, C updates his key as follow:
  $$C^j_x(w)=C^{j-1}_z(w),C^j_z(w)=C^{j-1}_x(w),$$
  and T updates his key as follow:
  $$T^j_x(w)=T^{j-1}_z(w),T^j_z(w)=T^{j-1}_x(w).$$
\item Let $G_j=P$ be the phase gate to be executed on wire $w\in c\cup t$. Using the fact that $PX=XZP$ (up to an irrelevant phase factor), we have:
 \begin{eqnarray}
  && P(X^{C^{j-1}_x(w)\oplus T^{j-1}_x(w)}Z^{C^{j-1}_z(w)\oplus T^{j-1}_z(w)})\nonumber\\
  &=& (X^{C^{j-1}_x(w)\oplus T^{j-1}_x(w)}Z^{C^{j-1}_x(w)\oplus C^{j-1}_z(w) \oplus T^{j-1}_x(w)\oplus T^{j-1}_z(w)})P.
  \end{eqnarray}
  Thus, C updates his key as follow:
  $$C^j_x(w)=C^{j-1}_x(w),C^j_z(w)=C^{j-1}_x(w)\oplus C^{j-1}_z(w),$$
  and T updates his key as follow:
  $$T^j_x(w)=T^{j-1}_x(w),T^j_z(w)=T^{j-1}_x(w)\oplus T^{j-1}_z(w).$$
\item Let $G_j=CNOT$ be the controlled-NOT gate to be executed on wires $w,w'\in c\cup t$. Assume that $w$ is the control while $w'$ is the target. The key-updating algorithm is different from that of Ref.\cite{dupuis2010}. The reason is as follow: the CNOT gate is performed by the third party (the server S), who possesses the two wires $w,w'$ while computing, thus the CNOT gate would not be  performed nonlocally. Using the fact that $\mathrm{CNOT}(X^jZ^k\otimes X^lZ^m)=(X^jZ^{k\oplus m}\otimes X^{j\oplus l}Z^m)\mathrm{CNOT}$, we have:
    \begin{eqnarray}
    && \mathrm{CNOT}(X^{C^{j-1}_x(w)\oplus T^{j-1}_x(w)}Z^{C^{j-1}_z(w)\oplus T^{j-1}_z(w)}\nonumber\\
    && \otimes X^{C^{j-1}_x(w')\oplus T^{j-1}_x(w')}Z^{C^{j-1}_z(w')\oplus T^{j-1}_z(w')}) \nonumber\\
    &=& (X^{C^{j-1}_x(w)\oplus T^{j-1}_x(w)}Z^{C^{j-1}_z(w)\oplus T^{j-1}_z(w)\oplus C^{j-1}_z(w')\oplus T^{j-1}_z(w')}\nonumber\\
    && \otimes X^{C^{j-1}_x(w)\oplus T^{j-1}_x(w)\oplus C^{j-1}_x(w')\oplus T^{j-1}_x(w')}Z^{C^{j-1}_z(w')\oplus T^{j-1}_z(w')})\mathrm{CNOT}.\nonumber
    \end{eqnarray}
    Thus, C updates his key as follow:
    \begin{eqnarray}
    && C^j_x(w)=C^{j-1}_x(w),C^j_z(w)=C^{j-1}_z(w)\oplus C^{j-1}_z(w'),\nonumber\\
    && C^j_x(w')=C^{j-1}_x(w)\oplus C^{j-1}_x(w'),C^j_z(w')= C^{j-1}_z(w'),
    \end{eqnarray}
    and T updates his key as follow:
    \begin{eqnarray}
    && T^j_x(w)=T^{j-1}_x(w),T^j_z(w)=T^{j-1}_z(w)\oplus T^{j-1}_z(w'),\nonumber\\
    && T^j_x(w')=T^{j-1}_x(w)\oplus T^{j-1}_x(w'),T^j_z(w')= T^{j-1}_z(w').
    \end{eqnarray}
\item Let $G_j=R$ be executed on wire $w\in c\cup t$. In the protocol, C and T apply quantum operators $X^rZ^{r'}P^{C^{j-1}_x(w)}$ and $X^sZ^{s'}P^{T^{j-1}_x(w)}$, respectively. According to the deduction in Ref.\cite{dupuis2010}, we have:
    \begin{eqnarray}
    && (X^sZ^{s'}P^{T^{j-1}_x(w)})(X^rZ^{r'}P^{C^{j-1}_x(w)})R(X^{C^{j-1}_x(w)\oplus T^{j-1}_x(w)}Z^{C^{j-1}_z(w)\oplus T^{j-1}_z(w)})=\nonumber\\
    && X^{s\oplus r\oplus C^{j-1}_x(w)\oplus T^{j-1}_x(w)}Z^{s'\oplus r'\oplus C^{j-1}_x(w)\oplus T^{j-1}_x(w)\oplus C^{j-1}_z(w)\oplus T^{j-1}_z(w)\oplus(r\oplus C^{j-1}_x(w))\cdot T^{j-1}_x(w)}R,\nonumber
    \end{eqnarray}
    where an irrelevant global phase factor is ignored. Because $r\oplus C^{j-1}_x(w)$ is known only by C and $T^{j-1}_x(w)$ is known only by T, an AND-BOX is called to compute two bits $\alpha,\beta$, such that $\alpha\oplus\beta=(r\oplus C^{j-1}_x(w))\cdot T^{j-1}_x(w)$. Assume C obtains $\alpha$ and T obtains $\beta$ after calling the AND-BOX. Thus, C updates his key as follow:
    \begin{equation}
    C^j_x(w)=r\oplus C^{j-1}_x(w),C^j_z(w)=r'\oplus\alpha\oplus C^{j-1}_z(w)\oplus C^{j-1}_x(w),
    \end{equation}
    and T updates his key as follow:
    \begin{equation}
    T^j_x(w)=s\oplus T^{j-1}_x(w),T^j_z(w)=s'\oplus\beta\oplus T^{j-1}_z(w)\oplus T^{j-1}_x(w).
    \end{equation}
\end{itemize}

\subsection{Analysis}
Protocol 2 is analyzed from the correctness and security (blindness and verifiability) in the following.

{\bf Theorem 4 (Correctness)}. Assume C, S and T follow the steps of Protocol 2, then the outcome is correct.

{\it Proof}. According to the key-updating algorithm, if the three parties follow the steps of Protocol 2, S carries out the UQC $C_\mathcal{U}$, and outputs the $n+m$-qubit state
\begin{equation}\label{eq3}
X^{C^k_x(c)\oplus T^k_x(c)}Z^{C^k_z(c)\oplus T^k_z(c)}(U_e|d\rangle)\otimes X^{C^k_x(t)\oplus T^k_x(t)}Z^{C^k_z(t)\oplus T^k_z(t)}(|e\rangle),
\end{equation}
where $|d\rangle$ is the data owned by C, $e\in\{0,1\}^m$ is the encoding owned by T, $U_e$ is the quantum transformation corresponding to the encoding $e$. S sends the former $n$ qubits to T, then T performs decryption transformation and gets $X^{C^k_x(c)}Z^{C^k_z(c)}(U_e|d\rangle)$, and sends it to C. Finally, C performs decryption transformation and gets the required result $U_e|d\rangle$. Thus, the outcome is correct.$\hfill{}\Box$

{\bf Theorem 5 (Blindness)}. Protocol 2 is blind.

{\it Proof}. In Protocol 2, C and T encrypt their privacy (the data and encoding) with quantum one-time pad before sending them, and their initial keys are selected locally by themselves. So no information about their initial keys is leaked. During the interaction of the protocol, they update their keys by locally performing a computing on their keys, except the execution of $R$ gate. Whenever a $R$ gate is executed, the key-updating algorithm needs to call an AND-BOX once. Thus, whether their keys is leaked depends on the security of the AND-BOX. Given an ideal AND-BOX, the information of keys (including every key that are obtained during the key-updating process) can be kept secret perfectly. Because the privacy is protected by the quantum one-time pad and the keys are not leaked, the privacy (the data, encoding and result) of C and T are protected perfectly. Thus, Protocol 2 is blind.$\hfill{}\Box$

Similar to Protocol 1, Protocol 2 does not satisfy strong verifiability. It only satisfies weak verifiability. How to transform it into a TBQC protocol with strong verifiability? This problem will be discussed in the next section.

{\bf Theorem 6}. Protocol 2 can defeat collusive attack.

{\it Proof}. In Protocol 2, all the privacy of T and C have been encrypted with quantum one-time pad when the protocol begins. During the whole process of computation, the privacy are in the state of being encrypted, and the decryption key is unknown by the others. For example, in the final step, the privacy are in the state shown by Eq.(\ref{eq3}), and the two decryption keys are $(C^k_x(c)\oplus T^k_x(c),C^k_z(c)\oplus T^k_z(c))$ and $(C^k_x(t)\oplus T^k_x(t),C^k_z(t)\oplus T^k_z(t))$, respectively. It can be seen that, for each key, half information is owned by C, and the other half is owned by T, and neither of them can decrypt it alone. Then, if S and C cooperate, they cannot attack the privacy of T. If S and T cooperate, they cannot attack the privacy of C. Thus, Protocol 2 can defeat collusive attack.$\hfill{}\Box$

\section{Discussions}
Both the TBQC protocols 1 and 2 satisfy the weak verifiability, that means the check of the result depends on the problem itself. How to design a TBQC protocol, such that it has strong verifiability? For this problem, a possible way is: firstly performing quantum authentication on the quantum data, and then performing TBQC protocol 2. Concretely, the TBQC protocol 2 can be improved using the technique proposed in Ref.\cite{dupuis2012}, where clifford-based quantum authentication code \cite{aharonov2008} is used to authenticate the result. The details will be given in the future.

Though the model presented here involves three participants, when two of them cooperate completely, it would degenerate to the BQC model that are usually studied. Thus, the TBQC model can be regarded as an extension of BQC model. If C and T cooperate completely, it corresponds to the BQC model, where the Client owns the data and algorithm and delegates his computational task to the Server. If T and S cooperate completely, it corresponds to the BQC model, where the Server provides the computational resources and algorithm, and the algorithm is private by the Client, see Ref.\cite{rohde2012}. If C and S cooperate completely, it corresponds to the BQC model, where one party has data and enough computational resources and the other party owns the algorithms, and the latter provides the algorithm to the former in a blind way. This case is worth for future research.

\section{Conclusions}
This paper extends the BQC model, and proposes a TBQC model. The TBQC model is a special case of three-party secure computation, where the three parties play the different roles. This case exists widely and is worth to study. Here, two concrete protocols are constructed. The first protocol cannot defeat collusive attack. The second one is constructed based on UQC. It is proved that it is correct, blind, and can defeat collusive attack.


\begin{thebibliography}{00}
%\softraggedright
%\itemsep=-4pt plus.2pt minus.2pt  %% sets the vertical space between items
%\small
\bibitem{childs2005}
Childs, A. M. (2005). ``Secure assisted quantum computation." Quantum Information \& Computation 5(6): 456-466.

\bibitem{vedral2012}
Vedral, V. (2012). ``Moving Beyond Trust in Quantum Computing." Science 335(6066): 294-295.

\bibitem{aharonov2008}
Aharonov, D., M. Ben-Or, et al. (2008). ``Interactive proofs for quantum computations." arXiv:0810.5375.

\bibitem{arrighi2006}
Arrighi, P. and L. Salvail (2006). ``Blind quantum computation." International Journal of Quantum Information 4(5): 883-898.

\bibitem{broadbent2009}
Broadbent, A., J. Fitzsimons, et al. (2009). Universal blind quantum computation. Foundations of Computer Science, 2009.

\bibitem{sueki2012}
Sueki, T., T. Koshiba, et al. (2012). ``Ancilla-Driven Universal Blind Quantum Computation." arXiv:1210.7450.

\bibitem{morimae2010}
Morimae, T., V. Dunjko, et al. (2010). ``Ground state blind quantum computation on AKLT state." arXiv:1009.3486.

\bibitem{morimae2012a}
Morimae, T. and K. Fujii (2012). ``Blind topological measurement-based quantum computation." Nature Communications 3: 1036.

\bibitem{morimae2012b}
Morimae, T. (2012). ``Continuous-variable blind quantum computation." Physical Review Letters 109(23): 230502.

\bibitem{dunjko2012}
Dunjko, V., E. Kashefi, et al. (2012). ``Blind quantum computing with weak coherent pulses." Physical Review Letters 108(20): 200502.

\bibitem{rohde2012}
Rohde, P. P., J. F. Fitzsimons, et al. (2012). ``Quantum Walks with Encrypted Data." Physical Review Letters 109(15): 150501.

\bibitem{fisher2013}
Fisher, K., A. Broadbent, et al. (2013). ``Quantum computing on encrypted data." arXiv: 1309.2586.

\bibitem{barz2012}
Barz, S., E. Kashefi, et al. (2012). ``Demonstration of Blind Quantum Computing." Science 335(6066): 303-308.

\bibitem{barz2013}
Barz, S., J. F. Fitzsimons, et al. (2013). ``Experimental verification of quantum computation." Nature Physics 9:
727-731.

\bibitem{fitzsimons2012}
Fitzsimons, J. F. and E. Kashefi (2012). ``Unconditionally verifiable blind computation." arXiv:1203.5217.

\bibitem{bera2010}
Bera, D., S. Fenner, et al. (2010). ``Efficient universal quantum circuits." Quantum Information \& Computation 10(1): 16-28.

\bibitem{dupuis2010}
Dupuis, F., J. B. Nielsen, et al. (2010). Secure two-party quantum evaluation of unitaries against specious adversaries. Crypto 2010.

\bibitem{liang2011}
Liang, M. and L. Yang (2011). ``Universal quantum circuit of near-trivial transformations." SCIENCE CHINA Physics, Mechanics \& Astronomy 54(10): 1819-1827.

\bibitem{bernstein1997}
Bernstein, E. and U. Vazirani (1997). ``Quantum complexity theory." SIAM Journal on Computing 26(5): 1411-1473.

\bibitem{dupuis2012}
Dupuis, F., J. B. Nielsen, et al. (2012). Actively secure two-party evaluation of any quantum operation. Crypto 2012.



\end{thebibliography}
\end{document}